\documentclass
[aps,prb,twocolumn,floatfix,english,showpacs,10pt,superscriptaddress]{revtex4-2}%
\usepackage{graphicx}
\usepackage{booktabs}
\usepackage{amsmath}
\usepackage{physics}
\usepackage{amssymb}
\usepackage{colordvi}
\usepackage{verbatim}
\usepackage{xcolor}
\usepackage{mathrsfs}
\usepackage{epsfig}
\usepackage{lipsum}
\usepackage{amsfonts}
\usepackage{makecell}
\usepackage[unicode=true, breaklinks=false, pdfborder={0 0 1}, backref=false,
colorlinks=true, linkcolor=blue, urlcolor=blue, citecolor=blue]{hyperref}%
\providecommand{\U}[1]{\protect\rule{.1in}{.1in}}
\setcitestyle{numbers,square}

\begin{document}

\title{Spin-orbit locking of magnons with localized microwave fields}

\author{Chengyuan Cai}
\thanks{These authors contributed equally to this work.}
\affiliation{School of Physics, Huazhong University of Science and Technology, Wuhan 430074, China}

\author{Zubiao Zhang}
\thanks{These authors contributed equally to this work.}
\affiliation{School of Physics, Huazhong University of Science and Technology, Wuhan 430074, China}

\author{Ji Zou}
\affiliation{Department of Physics, University of Basel, Klingelbergstrasse 82, 4056 Basel, Switzerland}

\author{Gerrit E. W. Bauer}

\affiliation{Kavli Institute for Theoretical Sciences, University of the Chinese Academy of Sciences, Beijing 100190, China}
\affiliation{WPI-AIMR and Institute for Materials Research and CSRN, Tohoku University, Sendai 980-8577, Japan}

\author{Tao Yu}
\email{taoyuphy@hust.edu.cn}
\affiliation{School of Physics, Huazhong University of Science and Technology, Wuhan 430074, China}

\date{\today}

\begin{abstract}
We address the photonic spin-orbit coupling known from nano-optics and plasmonics in the microwave regime. The spin $\mathbf{S}$ and momentum $\mathbf{q}$ of microwaves emitted by an excited magnetic particle are locked by $\mathbf{q}\cdot\mathbf{S}=0$ with a fixed chirality $\hat{\mathbf{n}}\cdot(\hat{\bf S}\times\hat{\bf q})=1$ when evanescent along $\hat{\mathbf{n}}\perp {\bf q}$. This field excites magnons in a nearby magnetic film in the form of directional beams that rotate with the magnetization direction. The exchange of these magnons between two distant nanomagnets leads to a highly tunable strong coupling and entangles their excited states.

\end{abstract}
\maketitle

\section{Introduction}

Photonic spin-orbit coupling (SOC) in nano-optics~\cite{optic1,optic2,Lodahl} and plasmonics~\cite{Bliokh1,plasmonic} refers to the locking of photon ``spin" and wave vector that allows, for example, unidirectional routing of photons and causes the photon quantum spin Hall effect.

The field of magnonics aims for the excitation, detection, and control of magnons, the quanta of spin wave excitations of magnetic order~\cite{Lenk1, Chumak1,Grundler1,Demidov1,Brataas1,Barman1,Flebus1}, generating memory functionality and logic circuits~\cite{magnon_memory,Dirk_ACS}. Magnons emit stray magnetic fields at gigahertz to terahertz frequencies that may feed microwave cavities and waveguides and interact with metals, superconductors, NV-centers in diamond, and other magnets.
Single magnons can be generated and manipulated by the coupling to superconducting qubits~\cite{quanta,Quantum11,Tabuchi1,Quirion1}. Magnons at surfaces of bulk magnets~\cite{Damon11,Walker11} and their stray fields in thin films ~\cite{chirality2} are chiral in the sense that their propagation normal to the magnetization is unidirectional, facilitating directional control of classical and quantum information flow~\cite{Application}. Mediated by magnetic stray fields magnons couple chirally with photons~\cite{magnon1,Zhu1,coupling1,Zhang1,Bimu,Fuli,One_way_steering}, other magnons~\cite{gamma,Liu1,Shiota1,Ishibashi1,Szulc1,Au1,Wang1,ZouPRL2024},
electrons~\cite{noncontact1}, Cooper pairs~\cite{pair1,pair2}, phonons~\cite{surface1,Xu1,Sasaki1,Tateno1,Shah1,Bas1}, and qubits~\cite{Ren1}. 
However, theoretical~\cite{magnon1,gamma,Au1,noncontact1,pair1,surface1,Ren1,Fuli,One_way_steering} and experimental~\cite{Xu1,pair2,Liu1,Shiota1,Ishibashi1,Szulc1,Wang1,Zhu1,Zhang1,coupling1,Bimu,Sasaki1,Tateno1,Shah1, Bas1} studies of magnon chirality have been limited to configurations with one spatial degree of freedom.

\begin{figure}[tbh]
\centering
\includegraphics[width=0.48\textwidth]{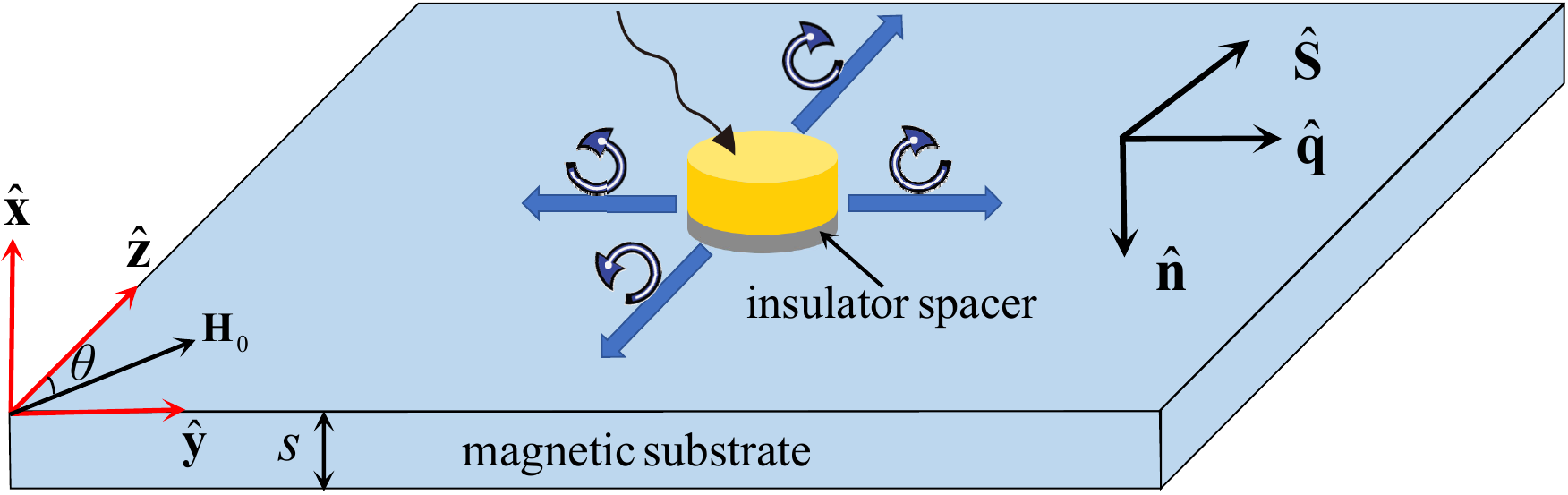}
\caption{Locking of microwave spin \textbf{S} (the curled arrows) and momentum \textbf{q} (the blue arrows) of the dynamic stray fields of a magnetic dot, governed by the chirality index $\hat{\bf n}\cdot({\hat{\bf S}\times \hat{\bf q}})=1$ that excites a nearby magnetic film of thickness \textit{s} into preferred directions. The stray field is evanescent along the normal $\hat{\bf n}$, and the hats indicate unit vectors.}
\label{model}
\end{figure}

Here, we unify the notion of photonic SOC at optical frequencies with the chiral coupling of magnets in the gigahertz regime by demonstrating that the spin $\mathbf{S}$ and
momentum $\mathbf{q}$-locked ac stray magnetic fields emitted by a point-like source in two dimensions results in a photonic (Rashba-like)
SOC $\mathbf{q}\cdot\mathbf{S}=0$ with chirality index
$\hat{\mathbf{n}}\cdot(\hat{\mathbf{S}}\times\hat{\mathbf{q}})=1$ (see
Fig.~\ref{model}). Here, hats indicate unit vectors, and the field must be evanescent normal to the plane of propagation  $\hat{\mathbf{n}}\perp {\bf q}$.
Moreover, the emitted radiation field is anisotropic, which allows the routing of magnons in extended ferromagnetic films when injected by nanomagnets under ferromagnetic resonance (FMR). The interaction of \textit{two} magnets by exchanging magnons in a magnetic substrate is strong and depends on the equilibrium magnetization directions. An on-chip distant entanglement ~\cite{Hill1,concurrencePRL} is, therefore, tunable without the reduced magnetic quality associated with structuring the films.

In Sec.~\ref{Sec_II}, we show that evanescent vector fields are always right-handed. We apply this general principle to the stray magnetic fields of excited nanomagnets in Sec~\ref{Section_III}. Sections~\ref{Section_IV} and \ref{Section_V} address two applications: routing the excited magnons in a nearby magnetic film and steering the interaction/entanglement between two nanomagnets.  Section~\ref{Section_VI} contains a discussion of the results, and technical details may be found in the Appendix.

\section{Fixed chirality of evanescent vector fields}

\label{Sec_II}

A fundamental property of
a vector field $\mathbf{V}(\mathbf{r},t)$ is its intrinsic angular momentum (or spin) distribution
$\mathbf{S}(\mathbf{r},t)\propto\mathrm{Im}\left(  \mathbf{V}^{\ast}\times\mathbf{V}\right)$~\cite{Bliokh1,chirality2}. For electric \textbf{E} and magnetic \textbf{H} fields in vacuum, ${\bf S}={\rm Im}(\varepsilon_0{\bf E}^*\times{\bf E}+\mu_0{\bf H}^*\times{\bf H})/(4\Omega)$, where $\Omega$ is the frequency and $\mu_0/\varepsilon_0$ is the vacuum permeability/permittivity. 
The spin of a Fourier component $\mathbf{q}$ is \textquotedblleft transverse\textquotedblright\ (\textquotedblleft
longitudinal\textquotedblright) when $\mathbf{S\bot q}$ $\left(\mathbf{S}\Vert\mathbf{q}\right)$.
Here we focus on the transverse
spin of plane wave modes that propagate in two dimensions but are evanescent along
a third field decay direction $\hat{\mathbf{n}}$. We measure the chirality by the index $\mathcal{C}_{\mathbf{q}}=\hat{\mathbf{n}}\cdot(\hat{\mathbf{S}}\times\hat{\mathbf{q}})$~\cite{Bliokh1,chirality2,chirality_3}, which is ``right-handed" when $0<\mathcal{C}_{\mathbf{q}}\leq1$, and ``left-handed" when $-1\leq\mathcal{C}_{\mathbf{q}}<0$. According to the Helmholtz theorem, any vector field can be decomposed into rotation- and divergence-free components. When there are no sources, the field is purely rotational.
We choose a coordinate system in which the vector field $\mathbf{V}(\mathbf{r},t)={\pmb {\cal V}}e^{i(q_{y}y+q_{z}z)+\sqrt{q_{y}%
^{2}+q_{z}^{2}}x}e^{-i\Omega t}$  propagates in the $y$-$z$ plane with wave
vector $\mathbf{q}=q_{y}\hat{\mathbf{y}}+q_{z}\hat{\mathbf{z}}$ and frequency $\Omega$, and is evanescent in the negative
half-space  $x<0$.
The amplitudes  ${\pmb {\cal V}}=(1,\mathcal{V}_{y}e^{i\phi_{y}},\mathcal{V}_{z}e^{i\phi_{z}})$ with
$\mathrm{Im}\mathcal{V}_{y,z}=0$ and phases $\{\phi_{y},\phi_{z}\}\in
\lbrack0,2\pi)$. 
In the absence of sources $\nabla\cdot\mathbf{V}(\mathbf{r},t)=0$  or 
\begin{align}
q_{y}\mathcal{V}_{y}\cos\phi_{y}+q_{z}\mathcal{V}_{z}\cos\phi_{z} &
=0,\nonumber\\
q_{y}\mathcal{V}_{y}\sin\phi_{y}+q_{z}\mathcal{V}_{z}\sin\phi_{z} &
=\sqrt{q_{y}^{2}+q_{z}^{2}}.
\label{divergence_free}
\end{align}
The spin
$\mathbf{S}(\mathbf{r},t)\propto\mathrm{Im}\left[  \mathbf{V}(\mathbf{r}%
,t)^{\ast}\times\mathbf{V}(\mathbf{r},t)\right]  $ has the components~\cite{Bliokh1}
\begin{align}
S_{x}(x,{q_{y}},q_{z}) &  \propto-2\mathcal{V}_{y}\mathcal{V}_{z}\sin(\phi
_{y}-\phi_{z})e^{2\sqrt{q_{y}^{2}+q_{z}^{2}}x},\nonumber\\
S_{y}(x,{q_{y}},q_{z}) &  \propto-2\mathcal{V}_{z}\sin{\phi_{z}}%
e^{2\sqrt{q_{y}^{2}+q_{z}^{2}}x},\nonumber\\
S_{z}(x,{q_{y}},q_{z}) &  \propto2\mathcal{V}_{y}\sin{\phi_{y}}e^{2\sqrt
{q_{y}^{2}+q_{z}^{2}}x}.
\label{spin_general}
\end{align}
Focusing on transverse vector fields,  
\begin{equation}
\mathbf{q}\cdot\mathbf{S}\propto2(q_{z}\mathcal{V}_{y}\sin{\phi_{y}}-q_{y}\mathcal{V}_{z}\sin{\phi_{z}})e^{2\sqrt{q_{y}^{2}+q_{z}^{2}}
x}=0.
\label{chiral}
\end{equation}
The modulus $\left\vert{\bf S}\right\vert$ depends on the field amplitudes. 
Equations~(\ref{divergence_free}) and (\ref{chiral}) fix the chirality index
$\mathcal{C}_{\bf q}\equiv (-\hat{\mathbf{x}})\cdot(\hat{\mathbf{S}}\times\hat{\mathbf{q}})$ as follows.
Except for the singular values $\phi_{y}\neq
\{0,\pi/2,\pi,{3\pi}/2\}$,
\begin{equation}
\frac{\mathcal{V}_{y}}{\mathcal{V}_{z}}=-\frac{q_{z}\cos\phi_{z}}{q_{y}%
\cos\phi_{y}}=\frac{q_{y}\sin{\phi_{z}}}{q_{z}\sin\phi_{y}}\label{free1}
\end{equation}
leads to $q_{z}^{2}\tan{\phi_{y}}+q_{y}^{2}\tan{\phi_{z}}=0$ and by substitution into
Eq.~(\ref{divergence_free}) $\mathcal{V}_{y}e^{i\phi_{y}}=q_{y}e^{i\phi
_{y}}/\left(  \sqrt{q_{y}^{2}+q_{z}^{2}}\sin\phi_{y}\right)  $ and
$\mathcal{V}_{z}e^{i\phi_{z}}=q_{z}e^{i\phi_{z}}/\left(  \sqrt{q_{y}^{2}
+q_{z}^{2}}\sin\phi_{z}\right)$,
such that
\begin{equation}
\mathcal{C}_{\mathbf{q}}=\left[  {\frac{q_{y}^{2}q_{z}^{2}\sin^{2}(\phi
_{y}-\phi_{z})}{(q_{y}^{2}+q_{z}^{2})^{2}\sin^{2}\phi_{y}\sin^{2}\phi_{z}}+1}\right]^{-1/2}>0.
\label{vector_field}
\end{equation}
Hence, any source-free evanescent vector field with transverse spin must be \textquotedblleft
right-handed\textquotedblright, implying the presence of a geometric SOC. 
Because $0<\left\vert \mathcal{C}_{\mathbf{q}%
}\right\vert \leq1,$ the spin cannot be perpendicular to the propagation plane. When the
spin lies in the plane with ${S}_{x}\rightarrow0$, the three vectors
$\{-\hat{\mathbf{x}},\hat{\mathbf{S}},\hat{\mathbf{q}}\}$ are at right angles, maximizing $\mathcal{C}_{\mathbf{q}}=1$.
When $\phi_{y}=\phi_{z}\rightarrow\{\pi/2,{3\pi}/2\}$,
$\mathcal{V}_{y}e^{i\phi_{y}}=iq_{y}/\sqrt{q_{y}^{2}+q_{z}^{2}}$,
$\mathcal{V}_{z}e^{i\phi_{z}}=iq_{z}/\sqrt{q_{y}^{2}+q_{z}^{2}}$, and
$\mathcal{V}_{y}/\mathcal{V}_{z}=\pm q_{y}/q_{z}$. Finally, for $\phi_{y}=0$
or $\pi$, $\phi_{z}=0$ or $\pi$,  $\sqrt{q_{y}^{2}+q_{z}^{2}}=0$, meaning that the field is standing. The precise value of \(\mathcal{C}_{\mathbf{q}}\) depends on the physical system and geometry under consideration.

Magnons are chiral by the dipolar coupling with other magnons and quasiparticles~\cite{magnon1,gamma,Au1,noncontact1,pair1,surface1,Ren1,Fuli,One_way_steering,Xu1,pair2,Liu1,Shiota1,Ishibashi1,Szulc1,Wang1,Zhu1,Zhang1,coupling1,Bimu,Sasaki1,Tateno1,Shah1, Bas1}.  Previous studies focus on configurations that can be reduced to one spatial dimension, such as long striplines on a thin film.  This prevented a full appreciation of the photonic spin-orbit coupling in magnonics. Here, we illustrate this point at the hands of magnetic devices that are effectively two-dimensional.

\section{Stray fields with maximal chirality}

\label{Section_III}

We illustrate the chirality of stray magnetic fields at the hand of a point magnetic moment $\tilde{\bf m}(t)$, noting that the finite size of the source can easily be incorporated by proper form factors (see below). The arguments equally apply to the electric stray fields of an electric dipole moment with transverse dynamics ${\bf p}(t)=p{\hat{\bf p}}(t)$~\cite{ferron}.

\subsection{Isotropic case}
\label{isotropic}

An in-plane magnetic field ${\bf H}_{0}$ at an angle ${\theta}$ with the
$\hat{\mathbf{z}}$-axis controls the dot magnetization $\tilde{\bf M}_{s}$. Here we first disregard the in-plane anisotropy by assuming $\tilde{\bf M}_{s}\parallel {\bf H}_{0}$ for simplicity and refer to the anisotropic situation in Sec.~\ref{anisotropic} below for a more general treatment. For convenience, we introduce a local $\{\tilde{x},\tilde{y},\tilde{z}\}$-reference frame with $\tilde{\mathbf{z}}\parallel \tilde{\bf M}_s$ and $\tilde{\mathbf{x}}\parallel\hat{\bf x}
\parallel\hat{\bf n}$. The dynamic magnetization of a weakly excited source 
\[\tilde{\bf M}(\mathbf{r}, t)=\delta({\mathbf{r}})\tilde{\bf m}(t)=\delta({\mathbf{r}})(\delta \tilde{m}e^{-i\Omega t},i\xi^{2}\delta
\tilde{m}e^{-i\Omega t},\tilde{m}_{s})^{T},
\]
where $\tilde{m}_s$ is the saturation magnetic moment,  $\delta \tilde{m}\ll\tilde{m}_{s}$
are  small transverse fluctuations, and $\xi^2>0$ is the ellipticity. By rotation to the laboratory $\{x,y,z\}$-frame, the components of the magnetic moment
$\tilde{m}_{x}=\delta \tilde{m} e^{-i\Omega t}$, $\tilde{m}_{y}=i\xi^{2}\delta \tilde{m} e^{-i\Omega t}\cos{\theta}+\tilde{m}_{s}%
\sin{\theta}$, $\tilde{m}_{z}=-i\xi^{2}\delta \tilde{m} e^{-i\Omega t}\sin{\theta}+\tilde{m}_{s}\cos{\theta}$.
The dipolar field follows from Coulomb's law
\begin{align}
    h_{\beta}(\mathbf{r},t)&=\frac{1}{4\pi}\partial_{\beta}\int_{-\infty}^{\infty}d{\mathbf{r}}^{\prime}\frac{\partial^{\prime}_{\alpha
}\tilde{M}_{\alpha}(\mathbf{r^{\prime}},t)}{|\mathbf{r}-\mathbf{r^{\prime}}|}\nonumber\\ &=\sum_{q_{y},q_{z}}e^{i(q_{y}y+q_{z}z)}h_{\beta}(x,q_{y},q_{z}
),
\label{Coulomb_law}
\end{align}
in the summation convention over repeated Cartesian indices $\{\alpha,\beta\}=\{x,y,z\}$. For $x<0$ the Fourier components
\begin{align}
h_{x}(x,q_{y},q_{z}) &  =\mathcal{F}_{q_{y},q_{z}} e^{\sqrt{q_{y}^{2}+q_{z}^{2}}x}/2,\nonumber\\
h_{y}(x,q_{y},q_{z}) &=\mathcal{F}_{q_{y},q_{z}} e^{\sqrt{q_{y}^{2}+q_{z}^{2}}x}{iq_y}/(2\sqrt{q_{y}^{2}+q_{z}^{2}}),\nonumber\\
h_{z}(x,q_{y},q_{z}) &=\mathcal{F}_{q_{y},q_{z}} e^{\sqrt{q_{y}^{2}+q_{z}^{2}}x}{iq_z}/(2{\sqrt{q_{y}^{2}+q_{z}^{2}}}),
\label{field_small}
\end{align}
where $\mathcal{F}_{q_{y},q_{z}}=\tilde{m}_{x}\sqrt{q_{y}^{2}+q_{z}^{2}%
}+\tilde{m}_{y}iq_{y}+\tilde{m}_{z}iq_{z}$.
The near field obeys the locking relation $iq_{y}
h_{y}(x,q_{y},q_{z})+iq_{z}h_{z}(x,q_{y},q_{z})=-\sqrt{q_{y}^{2}+q_{z}^{2}
}h_{x}(x,q_{y},q_{z})$.

\begin{figure}[tbh]
\centering\includegraphics[width=0.48\textwidth,trim=0.6cm 0cm 0cm 0.1cm]{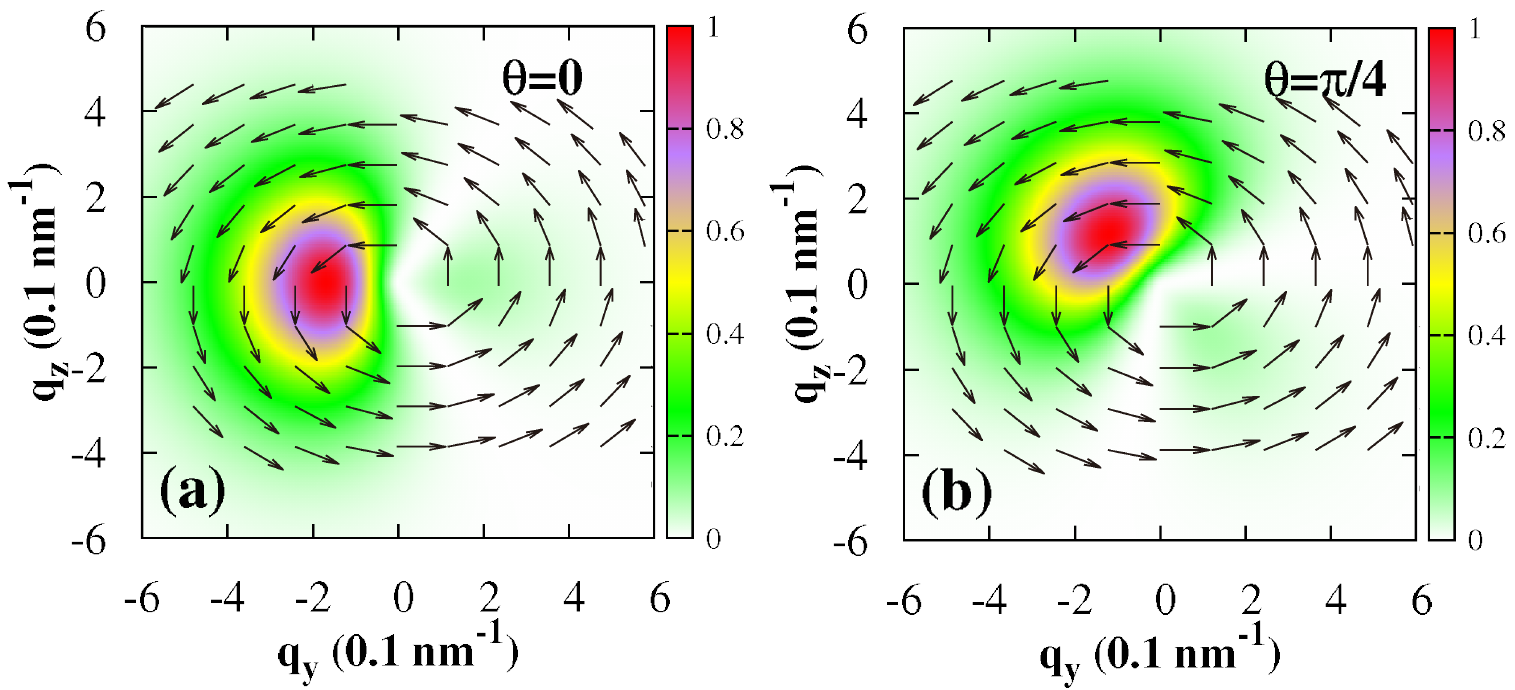}
\caption{Spin density $\mathbf{S}$ of the evanescent magnetic fields emitted by a
point source at a distance $x=-6~\mathrm{nm}$ as a function of in-plane wave numbers $q_{y}$ and $q_{z}$  and different directions of the in-plane applied magnetic field $\theta=\{0,\pi/4\}$. The arrows represent the direction and the colors of the modulus of ${\bf S}$. An analytic estimate of the wave number maximizing \(|{\bf S}|\) in polar coordinates is ${\bf q}_{\mathrm{max}}=(-1/x,\pi-\theta)$ and \(-1/x=0.17\)~nm\(^{-1}\), which  agrees with the numerical results. }
\label{spin_2}
\end{figure}

The spin density of the stray magnetic field \eqref{field_small}~\cite{Walker11,
Damon11, Kino1,Viktorov1,Bliokh1,chirality_3} 
\begin{align}
    \mathbf{S}(x,{q_{y}},q_{z})=\frac{\mu_0}{4\Omega}\mathrm{Im} \left[  \mathbf{h}^{\ast}
(x,q_{y},q_{z})\times\mathbf{h}(x,q_{y},q_{z})\right].
\label{Spin_density_definition}
\end{align}
In the lower half-space ($x<0$) 
\begin{align}
S_{x}(x,{q_{y}},q_{z}) & =0,\nonumber\\
S_{y}(x,{q_{y}},q_{z}) & = -\frac{\mu_0}{8\Omega}|\mathcal{F}_{q_{y},q_{z}}|^{2}q_{z} {e^{2\sqrt{q_{y}^{2}+q_{z}^{2}}x}}/{\sqrt{q_{y}%
^{2}+q_{z}^{2}}},\nonumber\\
S_{z}(x,{q_{y}},q_{z}) & =\frac{\mu_0}{8\Omega}|\mathcal{F}_{q_{y},q_{z}}|^{2} q_{y}{e^{2\sqrt{q_{y}^{2}+q_{z}^{2}}x}}/{\sqrt{q_{y}%
^{2}+q_{z}^{2}}}.
\label{spin}
\end{align}
This field is perfectly spin-momentum locked with $C_{\mathbf{q}}=1$ and therefore realizes Eq.~\eqref{vector_field} with fixed phase $\phi_{y}=\phi_{z}=\pi/2$ and spin
$\mathbf{S}$ lying in the propagation plane. 
The Rashba SOC  of free electrons obeys the same relation, \textit{viz.}
$\mathbf{q}\cdot\mathbf{S}=0$ with integer chirality $\hat{\mathbf{n}}\cdot(\hat{\mathbf{S}}\times\hat{\mathbf{q}})=1$.
In polar coordinates, the spin density ${\bf S}(x,{\bf q})$ is maximized by ${\bf q}=(-1/x,\pi-\theta)$. 
Figure~\ref{spin_2} shows plots of the spin density
$\mathbf{S}$ as a function of the wave numbers $q_{y}$ and $q_{z}$ for $\xi^2=3.3$ and the parameters in Fig.~\ref{excitation_11}(a)-(c) below. The chirality of the dipolar field is indeed always \textquotedblleft right-handed\textquotedblright.

\subsection{Anisotropic case}
\label{anisotropic}

\subsubsection{Cuboid shape anisotropy}

Here, we address the Kittel mode of cuboid nanomagnets of width $w$, length $l$, and thickness $d$ and in-plane magnetic field ${\bf H}_{0}$ applied at an angle $\theta$ with the $\hat{\mathbf{z}}$-axis, but disregard crystal anisotropies.
The equilibrium tilt angle $\tilde{\theta}$ of the saturation magnetization $\tilde{\mathbf{M}}_s$ minimizes the free energy density
\begin{align}
    F_m&=-\mu_0\tilde{M}_sH_0\cos(\theta-\tilde{\theta})\nonumber\\&+\frac{\mu_0}{2} N_{yy}\tilde{M}_s^2\sin^2{\tilde\theta}+\frac{\mu_0}{2} N_{zz}\tilde{M}_s^2\cos^2{\tilde\theta},
\end{align}
where  $N_{x x}\simeq {wl}/{(wl+wd+ld)}$, $N_{y y}\simeq {ld}/{(wl+wd+ld)}$, and $N_{z z}\simeq{wd}/{(wl+wd+ld)}$ are the demagnetization factors and $\tilde{M}_s=|\tilde{\bf M}_s|$.
At equilibrium $dF_m/d\tilde\theta=0$ and $d^2F_m/d\tilde\theta^2\textgreater0$, or
\begin{align}
    &H_0\sin(\theta-\tilde{\theta})+\frac{\tilde{M}_s}{2}(N_{zz}-N_{yy})\sin({2\tilde\theta})=0,\nonumber\\
    &H_0\cos(\theta-\tilde{\theta})-\tilde{M}_s(N_{zz}-N_{yy})\cos({2\tilde\theta})\textgreater 0.
\end{align}

The magnetization \(\tilde{\bf M}\) of the nanomagnets precesses around the effective magnetic field ${\bf H}_{\rm eff}$, which is the sum of the applied static field ${\bf H}_0$ and the demagnetization field ${\bf H}_d$. In the lab frame
\begin{align}
    {\bf H}_{\rm eff}={\bf H}_0+{\bf H}_d=\left(\begin{array}{ccc}
		& -N_{xx}{\tilde{M}}_{x} \\
		& -N_{yy}{\tilde{M}}_{y}+H_0\sin\tilde{\theta}\\ 
		& -N_{zz}{\tilde{M}}_{z}+H_0\cos\tilde{\theta}
	\end{array}\right).
 \label{block_field}
\end{align}
In the local $\{\tilde{x},\tilde{y},\tilde{z}\}$-coordinate system with $\tilde{\mathbf{z}}\parallel \tilde{\bf M}_s$ and $\tilde{\mathbf{x}}\parallel\hat{\bf x}$, the linearized Landau-Lifshitz equation simplifies to
\begin{align}
    &\partial {\tilde{M}}_{\tilde{x}}/{\partial t}=-\omega_1{\tilde{M}}_{\tilde{y}},\nonumber \\
&\partial {\tilde{M}}_{\tilde{y}}/{\partial t}=\omega_2 {\tilde{M}}_{\tilde{x}},
\label{block_LL}
\end{align}
where $\omega_1=\mu_0 \gamma[H_0 \cos\small(\theta-\tilde{\theta}\small)-\tilde{M}_s(\cos ^2 \tilde{\theta}-\sin ^2 \tilde{\theta})(N_{z z}-N_{y y})]$,  $\omega_2=\mu_0 \gamma[H_0 \cos \small(\theta-\tilde{\theta}\small)-\tilde{M}_s(N_{y y} \sin ^2 \tilde{\theta}+N_{z z} \cos ^2 \tilde{\theta}-N_{xx})]$, and $\gamma$ is the electron gyromagnetic ratio. Its solution leads to the resonance frequency
\begin{align}
\Omega=\sqrt{\omega_1\omega_2},
\end{align}
and an ellipticity
\begin{align}
\xi^2\equiv -i{\tilde{M}}_{\tilde{y}}/{\tilde{M}}_{\tilde{x}}=\sqrt{\omega_2/\omega_1}.
\label{ellipticity}
\end{align}

\begin{figure}[tbh]
\centering\includegraphics[width=0.48\textwidth,trim=0.6cm 0cm 0cm 0.1cm]{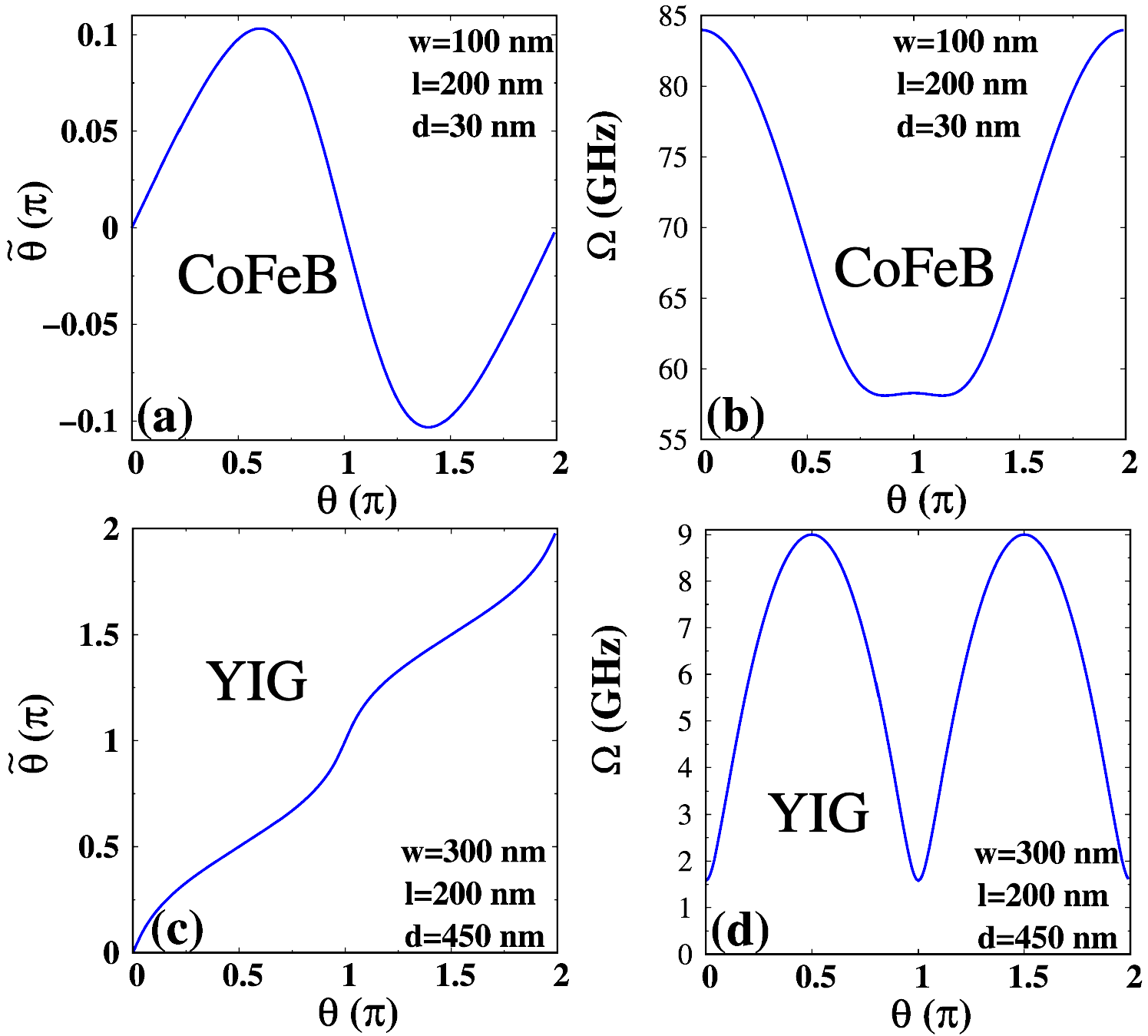}
\caption{Dependence of canting angle $\tilde\theta$ of the magnetization $\tilde{\bf M}_s$ and FMR frequency $\Omega$ on the direction $\theta$ of an applied field $\mu_{0}H_{0}=0.05$~T for cubic nanoparticle of CoFeB with $\mu_{0}\tilde{M}_{s}=1.6$~T [(a) and (b)] and YIG with $\mu_{0}\tilde{M}_{s}=0.177$~T [(c) and (d)].}
\label{magnetization_1}
\end{figure}

Figure~\ref{magnetization_1} plots the dependence of the direction $\tilde\theta$ of $\tilde{\bf M}_s$  and FMR frequency $\Omega$ on the direction $\theta$ of an applied field  $\mu_{0}H_{0}=0.05$~T  in typical materials.  $\tilde{\theta}\approx 0$ is pinned along the initial $\hat{\bf z}$-direction in the hard CoFeB magnet of width $w=100~\mathrm{nm}$, length $l=200~\mathrm{nm}$, thickness
$d=30~\mathrm{nm}$ [Fig.~\ref{magnetization_1}(a) and (b)]. The soft magnetization of a yttrium iron garnet (YIG) particle of dimensions $\{w,l,d\}=\{300,200,450\}$~nm can be almost freely rotated since  $\tilde{\theta}\approx \theta$ [Fig.~\ref{magnetization_1}(c) and (d)].

\subsubsection{Chirality of stray magnetic field}

The chirality of the stray fields of point magnetic sources persists when emitted by finite anisotropic nanomagnets. The Kittel mode dynamics still reads $\tilde{\bf M}(t)=(\delta \tilde{M}e^{-i\Omega t},i\xi^{2}\delta
\tilde{M}e^{-i\Omega t},\tilde{M}_{s})^{T}$, where $\delta \tilde{M}\ll\tilde{M}_{s}$
represents the transverse fluctuation of the magnetization and $\xi^{2}$ is the
ellipticity \eqref{ellipticity}.
By rotation to the laboratory $\{x,y,z\}$-frame
\begin{align}
\tilde{M}_{x} &  =\delta \tilde{M}e^{-i\Omega t},\nonumber\\
\tilde{M}_{y} &  =i\xi^{2}\delta \tilde{M}e^{-i\Omega t}\cos\tilde{\theta}+\tilde{M}_{s}%
\sin\tilde{\theta},\nonumber\\
\tilde{M}_{z} &  =-i\xi^{2}\delta \tilde{M}e^{-i\Omega t}\sin\tilde{\theta}+\tilde{M}%
_{s}\cos\tilde{\theta}.
\end{align}
The Fourier components for $x<0$ of the dynamic fields 
\begin{align}
h_{x}(x,q_{y},q_{z}) &  =\mathcal{F}_{q_{y},q_{z}}V_{q_{y},q_{z}}%
(x)\sqrt{q_{y}^{2}+q_{z}^{2}},\nonumber\\
h_{y}(x,q_{y},q_{z}) &  =\mathcal{F}_{q_{y},q_{z}}V_{q_{y},q_{z}}%
(x)iq_{y},\nonumber\\
h_{z}(x,q_{y},q_{z}) &  =\mathcal{F}_{q_{y},q_{z}}V_{q_{y},q_{z}}%
(x)iq_{z},\label{field_small}%
\end{align}
where $\mathcal{F}_{q_{y},q_{z}}=\tilde{M}_{x}\sqrt{q_{y}^{2}+q_{z}^{2}%
}+\tilde{M}_{y}iq_{y}+\tilde{M}_{z}iq_{z}$ and the form factor
\begin{align}
V_{q_{y},q_{z}}(x) &  =\frac{2}{({q_{y}^{2}}+{q_{z}^{2}})q_{y}q_{z}}\sin
(q_{y}\frac{w}{2})\sin(q_{z}\frac{l}{2})\nonumber\\&\times e^{\sqrt{q_{y}^{2}+q_{z}^{2}}x}\left(  1-e^{-\sqrt{q_{y}^{2}%
+q_{z}^{2}}d}\right),
\label{field4}%
\end{align}
peak around $\left\vert q_{y}\right\vert \sim
\pi/w$ and $|q_{y}|\sim\pi/l$ and obey the relation 
\begin{align}
q_{y}h_{y}(x,q_{y},q_{z})+q_{z}h_{z}(x,q_{y},q_{z})=i\sqrt{q_{y}^{2}+q_{z}^{2}%
}h_{x}(x,q_{y},q_{z}).
\end{align}

Their spin density~\eqref{Spin_density_definition} in the lower half-space $x<0$ reads
\begin{align}
S_{x}(x,{q_{y}},q_{z})&=0,\nonumber\\
S_{y}(x,{q_{y}},q_{z})&=-\frac{\mu_0}{2\Omega}|\mathcal{F}_{q_{y},q_{z}}|^{2}V_{q_{y},q_{z}}^{2}(x)\sqrt{q_{y}%
^{2}+q_{z}^{2}}q_{z},\nonumber\\
S_{z}(x,{q_{y}},q_{z})&=\frac{\mu_0}{2\Omega}|\mathcal{F}_{q_{y},q_{z}}|^{2}V_{q_{y},q_{z}}^{2}(x)\sqrt{q_{y}%
^{2}+q_{z}^{2}}q_{y}.
\end{align}
The field is still spin-momentum locked with maximal chirality index
$C_{\mathbf{q}}=1$.

\begin{figure}[tbh]
\centering\includegraphics[width=0.4\textwidth,trim=0.6cm 0cm 0cm 0.1cm]{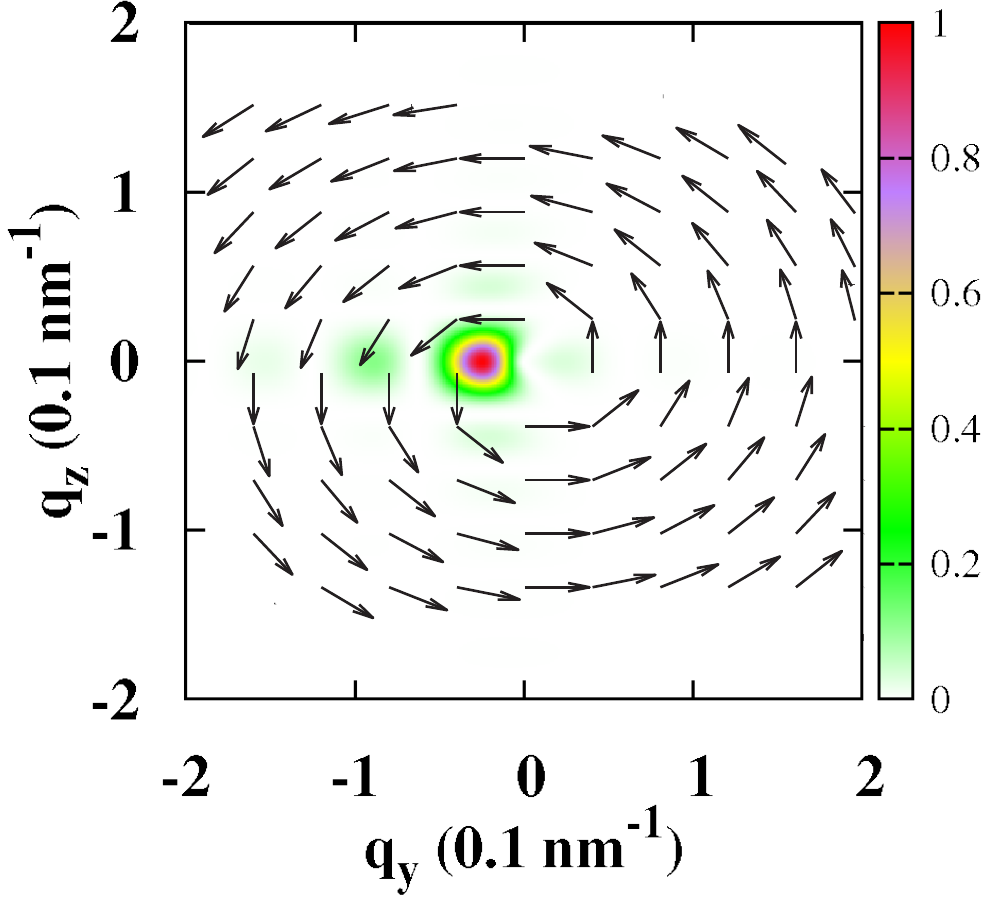}
\caption{Fourier components of the spin density $\mathbf{S}$ of the stray magnetic field emitted by a resonantly excited CoFeB particle at $x=-6~\mathrm{nm}$ (parameters are given in the text) for $\tilde{\bf M}_s\parallel \hat{\mathbf{z}}$, $\tilde{\theta}=0$. The arrows represent the direction and the background color of the modulus of $\mathbf{S}$.}
\label{spin_1}
\end{figure}

Figure~\ref{spin_1} plots the dependence of the spin density $\mathbf{S}$ on the wave numbers $q_{y}$ and $q_{z}$ for the CoFeB nanomagnet as in Fig.~\ref{magnetization_1}(a) and (b) and a magnetic field $\mu_{0}\textbf{H}_{0}= 0.05$~T \(\hat{\mathbf{z}}\), illustrating its
\textquotedblleft right-handedness\textquotedblright.  The results are qualitatively the same as those for the point magnet in Fig.~\ref{spin_2}. However, the form factor suppresses the higher momentum components, and the emitted beams are better focused.

\section{Steering magnons}

\label{Section_IV}

The spin-momentum locking of the stray field strongly affects the coupling to other quasi-particles, such as magnons in an underlying magnetic film.

\subsection{Isotropic case}

\label{isotropic_II}

We first consider an ultrathin and soft magnetic film of thickness $s$ excited by a magnetic disk of radius $w$ and thickness $d$ and turn to the effect of anisotropies in Sec.~\ref{anisotropic_II} below. An optional thin insulating spacer between the source and substrate suppresses the interfacial exchange interaction while leaving their long-ranged dipolar interaction unaffected~\cite{Chen1,Dirk_ACS}. Without anisotropies, the equilibrium magnetizations and the in-plane magnetic field ${H}_{0}$ applied at an angle $\theta$ are always parallel. 
In sufficiently thin films, the lowest perpendicular standing spin wave frequencies lie well below that of the higher subbands. Disregarding small dipolar corrections, the dispersion  $\omega(k)=\mu_{0}\gamma(H_{0}+\alpha_{\mathrm{ex}}M_{s}k^{2})$, where $\alpha_{\mathrm{ex}}$ is the exchange stiffness and $\gamma$ is the modulus of the electron gyromagnetic ratio.
Retaining the lowest term of the Holstein-Primakoff expansion~\cite{Kittel_book1,HP1} of the spin operator
$\hat{\mathbf{S}}$, the  magnetization operator $\hat{\mathbf{M}}=-\gamma
\hbar\hat{\mathbf{S}}$ of the film 
\begin{align}
\hat{M}_{y}(\pmb{\rho}) &  =M_{s}\sin{\theta}-\sqrt{2M_{s}\gamma\hbar}\sum_{\mathbf{k}}(
\mathcal{M}_{y}e^{i\mathbf{k}\cdot\pmb{\rho}}\hat{m}_{\mathbf{k}%
}+\mathrm{h.c.})  \cos{\theta},\nonumber\\
\hat{M}_{z}(\pmb{\rho}) &  =M_{s}\cos{\theta}+\sqrt{2M_{s}\gamma\hbar}\sum_{\mathbf{k}}(
\mathcal{M}_{y}e^{i\mathbf{k}\cdot\pmb{\rho}}\hat{m}_{\mathbf{k}%
}+\mathrm{h.c.})  \sin{\theta},\nonumber\\
\hat{M}_{x}(\pmb{\rho}) &  =-\sqrt{2M_{s}\gamma\hbar}\sum_{\mathbf{k}}\left(
\mathcal{M}_{x}e^{i\mathbf{k}\cdot\pmb{\rho}}\hat{m}_{\mathbf{k}%
}+\mathrm{h.c.}\right),
\label{film_lab}%
\end{align}
where the in-plane position vector $\pmb{\rho}=y\hat{\mathbf{y}}%
+z\hat{\mathbf{z}}$ and $\hat{m}_{\mathbf{{k}}}$ annihilates a magnon with
wave vector $\mathbf{{k}}$. For circular polarization
 $\mathcal{M}_{x}=-1/({2\sqrt{L_{y}L_{z}s}})$ and $\mathcal{M}_{y}=-{i}%
/({2\sqrt{L_{y}L_{z}s}})$~\cite{Sanchar_PRB1}. 
$\hat{m}_{\mathbf{k}}$ interacts with the stray
field of a Kittel magnon $\hat{\beta}$ of the nanomagnet by Zeeman
interaction~\cite{Kittel1}
\begin{align}
    \hat{H}_{c}&=-\mu_{0}\int_{-s}^{0}dx\int_{-\infty}^{\infty}dydz\hat{\mathbf{h}%
}(\mathbf{r})\cdot\hat{{\mathbf{M}}}(\mathbf{r})\nonumber\\
&=\sum_{\mathbf{k}}\hbar
g_{\mathbf{k}}\hat{m}_{\mathbf{k}}\hat{\beta}^{\dagger}+\mathrm{H.c.}, 
\end{align}
with coupling constant
\begin{align}
g_{\mathbf{k}} &  =-4\pi\mu_0\gamma w\sqrt{M_{s}\tilde{M}_{s}}\left(
1-e^{-kd}\right)  \left(  1-e^{-ks}\right)J_{1}(kw)  \nonumber\\
&  \times \frac{1}{k^4} \left(  \mathcal{M}_{x},\mathcal{M}_{y}\right)  \left(
\begin{array}
[c]{cc}%
k^{2} & -ik\kappa\\
-ik\kappa& -\kappa^2
\end{array}
\right)  \left(
\begin{array}
[c]{c}%
\tilde{\mathcal{M}}_{\tilde{x}}^{\ast}\\
\tilde{\mathcal{M}}_{\tilde{y}}^{\ast}%
\end{array}
\right),
\nonumber
\end{align}
where $J_1(x)$ is the first-order Bessel function of the first kind,
$\kappa= k_y\cos\theta- k_z\sin\theta$,  $\tilde{\mathcal{M}}_{\tilde{x}}=-1/(2\xi\sqrt{\pi w^2d})$ and $\tilde{\mathcal{M}}_{\tilde{y}}=-i\xi/(2\sqrt{\pi w^2d})$ are the amplitudes of Kittle modes in the nanomagnet, and the ellipticity $\xi^2=\sqrt{(H_0+(N_{\bot}-N_{||}){\tilde M}_s)/H_0}$ with demagnetization factors $N_{||}\simeq d/(2d+\sqrt{\pi}w)$ and $N_{\bot}\simeq \sqrt{\pi}w/(2d+\sqrt{\pi}w)$~\cite{disk}.

The associated quantum Langevin equation of motion~\cite{Gardiner,Clerk} $i{d\hat{m}_{\mathbf{k}}}/{dt}=(\omega_{k}-i\delta_{m})\hat
{m}_{\mathbf{k}}+g_{\mathbf{k}}{\hat{\beta}}$ and $i{d\hat{\beta}}/{dt}=(\Omega-i\delta_{\beta})\hat{\beta}+\sum_{\mathbf{k}}g_{\mathbf{k}}\hat{m}_{\mathbf{k}}$, in which $\Omega=\mu_0\gamma\sqrt{H_0(H_0+(N_{\bot}-N_{||}){\tilde M}_s)}$ lies in the continuum of $\omega_{k}$, $\delta_{\beta}=\tilde{\alpha}_{G}\Omega$, and $\delta_{m}=\alpha
_{G}\omega_{k}$ with $\tilde{\alpha}_{G}$ and $\alpha_{G}$ denoting,
respectively, the damping constants of the nanomagnets and film. Exciting the
nanomagnet resonantly by microwaves of frequency $\Omega$ to an amplitude $\langle{\hat{\beta}}(\omega)\rangle$, $\langle\hat{m}_{\mathbf{k}}(\omega)\rangle={g_{\mathbf{k}}}\langle{\hat{\beta}}(\omega)\rangle/({\Omega
-\omega_{k}+i\delta_{m}})$.
Substituting into Eq.~(\ref{film_lab}), we obtain the excited magnetization in
the film,
\begin{align}
&  \langle\hat{M}_{x}(\pmb{\rho})\rangle=-\sqrt{2M_{s}\gamma\hbar}\left(
\mathcal{M}_{x}G(\pmb{\rho})\langle\hat{\beta}\rangle+\mathrm{h.c.}\right)
,\nonumber\\
&  \langle\hat{M}_{y}(\pmb{\rho})\rangle=-\sqrt{2M_{s}\gamma\hbar}\left(
\mathcal{M}_{y}G(\pmb{\rho})\langle\hat{\beta}\rangle+\mathrm{h.c.}\right)
\cos{\theta},\nonumber\\
&  \langle\hat{M}_{z}(\pmb{\rho})\rangle=\sqrt{2M_{s}\gamma\hbar}\left(
\mathcal{M}_{y}G(\pmb{\rho})\langle\hat{\beta}\rangle+\mathrm{h.c.}\right)
\sin{\theta},\label{film_excite}%
\end{align}
where the Green function
\begin{align}
&G(\pmb{\rho})   =\sum_{\mathbf{k}}e^{i\mathbf{k}\cdot\pmb{\rho}}\frac{g_{\mathbf{k}}%
}{\Omega-\omega_{k}+i\delta_{m}}=-i\frac{L_{y}L_{z}}{4\pi}\nonumber\\
&  \times%
\begin{cases}
\int_{0}^{2\pi}d\varphi({k_{\Omega}}/{v_{k_{\Omega}}})g(k_{\Omega},\varphi), &
\rho=0\\
\int_{\phi-\pi/2}^{\phi+\pi/2}d\varphi({2k_{\Omega}}/{v_{k_{\Omega}}%
})g(k_{\Omega},\varphi)e^{iq_{\Omega}\rho\cos(\varphi-\phi)}, & \rho\neq0
\end{cases}.
\nonumber
\end{align}
Here $k_{\Omega}=\sqrt{(\Omega-\mu_{0}\gamma H_{0})/(\mu_{0}\gamma
\alpha_{\mathrm{ex}}M_{s})}$ is a resonant wave number, $q_{\Omega}=k_{\Omega}(1+i\alpha_G/2)$, and in polar
coordinates $\pmb{\rho}=\{\rho,\phi\}$.

\begin{figure}[tbh]
\centering
\includegraphics[width=1.0\linewidth]{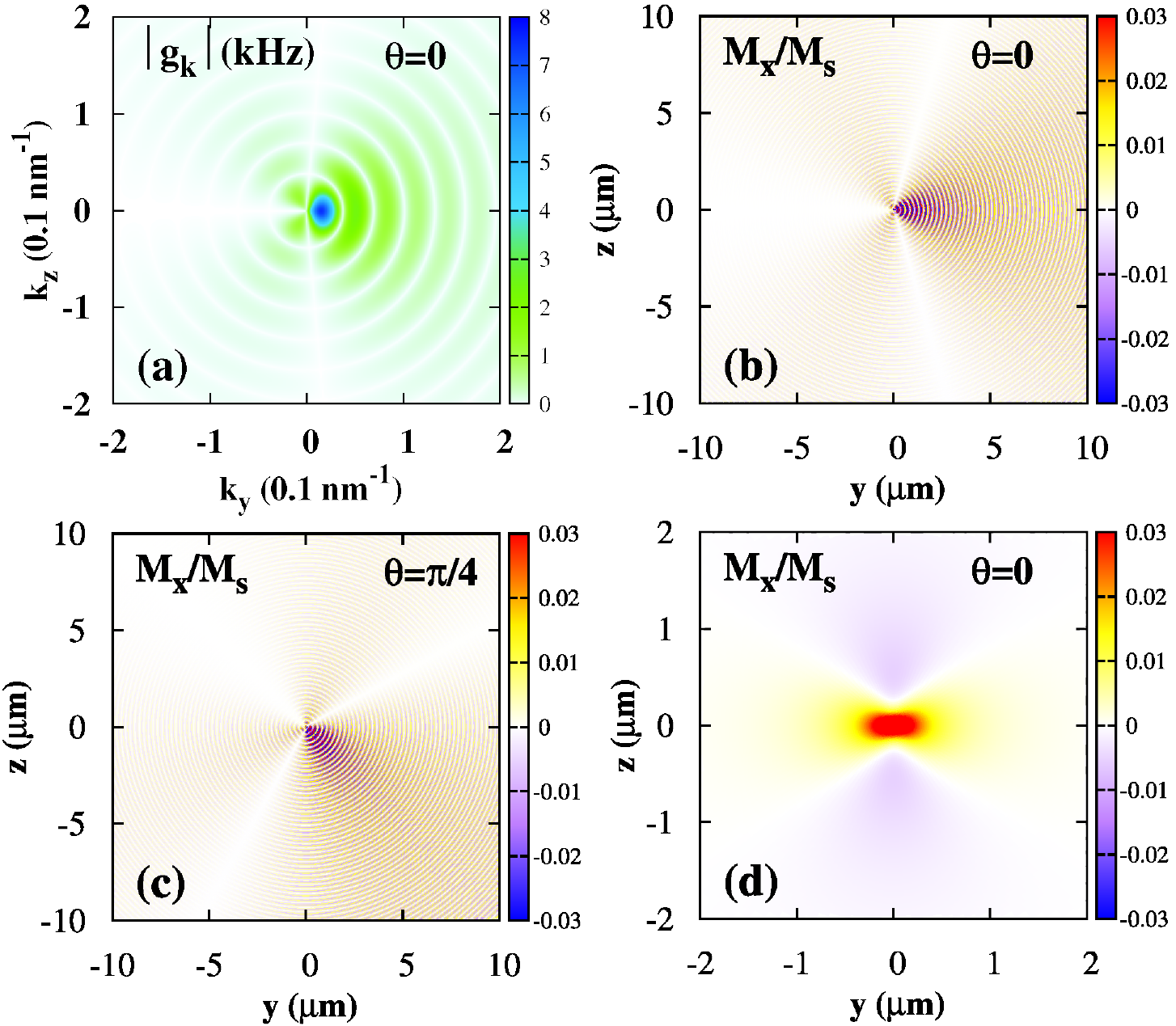} \hspace{0.2cm}
\caption{Magnetization dynamics of a thin magnetic film excited by a magnetic disc under FMR with frequency $\Omega$, \textit{cf}. Fig.~\ref{model}. (a) is the coupling constant $g_{\bf k}$.  (b) and (c) show the routing of resonantly excited spin waves with $\Omega>\omega_0$. Since disc and film are here magnetically soft, a variation of the magnetic field
 direction $\theta=\{0,\pi
/4\}$ rigidly rotates the magnetization distribution. (d) illustrates the magnetization dynamics under non-resonant microwave excitation  $\Omega<\omega_0$. 
For the  parameters of Fig.~\ref{angle_11}, $\Omega=\mu_0\gamma\sqrt{H_0(H_0+(N_{\bot}-N_{||}){\tilde M}_s)}=18.1$~GHz and $\omega_0=\mu_0\gamma H_0=18.2$~GHz. Other parameters are given in the text.}
\label{excitation_11}
\end{figure}

In Fig.~\ref{excitation_11}(a), we plot the momentum-dependent coupling constant $g_{\bf k}$, while Figs.~\ref{excitation_11}(b) and (c) show the
excited magnetization texture in a thin YIG film of $s=10~\mathrm{nm}$ directly below a CoFeB disk with dimensions $\{w,d\}=\{300,50\}~\mathrm{nm}$ and excitation amplitude
$\langle\hat{\beta}\rangle=1\times10^{6}$ or $\tilde{M}_{x}/\tilde{M}_{s}\approx0.03$, 
with $\mu_{0}{M}_{s}=0.177$~T~\cite{Wang1}, exchange stiffness
$\alpha_{\mathrm{ex}}=3\times10^{-16}~\mathrm{m}^{2}$~\cite{Wang1}, 
Gilbert damping constant ${\alpha}_{G}=10^{-4}$, $\gamma=1.82\times
10^{11}~\mathrm{s^{-1}\cdot T^{-1}}$, and  $\mu_{0}\tilde{M}_{s}%
=1.6$~T~\cite{Heigl1}. 
We find an anisotropic ``lighthouse" distribution of the emitted magnons, i.e., beams that can be steered by the direction of YIG's (soft) equilibrium
magnetization owing to the photonic SOC (Fig.~\ref{spin_2}). The interference features caused here by the local source should not be confused with spin-wave caustics excited by microwave striplines that reflect the anisotropy of the
spin-wave dispersion in thicker films~\cite{IBertellisci}. Diamond NV-center microscopy is the method of choice to confirm our predictions~\cite{IBertellisci}.

Figure~\ref{excitation_11}(d) shows the film magnetization dynamics when the field frequency lies \textit{below} the spin-wave continuum ($\Omega < \omega_0$) for the parameters specified in Fig.~\ref{angle_11} below. We can understand the suppressed chirality in terms of the symmetry of the Green function
    $G(-\pmb{\rho})=\sum_{\mathbf{k}}e^{-i\mathbf{k}\cdot\pmb{\rho}}{g_{\mathbf{k}}}/({\Omega-\omega_{k}+i\delta_{m}})\approx G(\pmb{\rho})^*$
since the intrinsic $\delta_m$ in the denominator is small. Consequently 
$\langle\hat{M}_{x}(\pmb{\rho})\rangle=-\sqrt{2M_{s}\gamma\hbar}\left(
\mathcal{M}_{x}G(\pmb{\rho})\langle\hat{\beta}\rangle+\mathrm{h.c.}\right)=\langle\hat{M}_{x}(-\pmb{\rho})\rangle$.
A larger detuning decreases the amplitude and exponential decay length of the coherently excited spin waves.

\subsection{Anisotropic case}

\label{anisotropic_II}

\begin{figure}[tbh]
\centering\includegraphics[width=0.48\textwidth,trim=0.6cm 0cm 0cm 0.1cm]{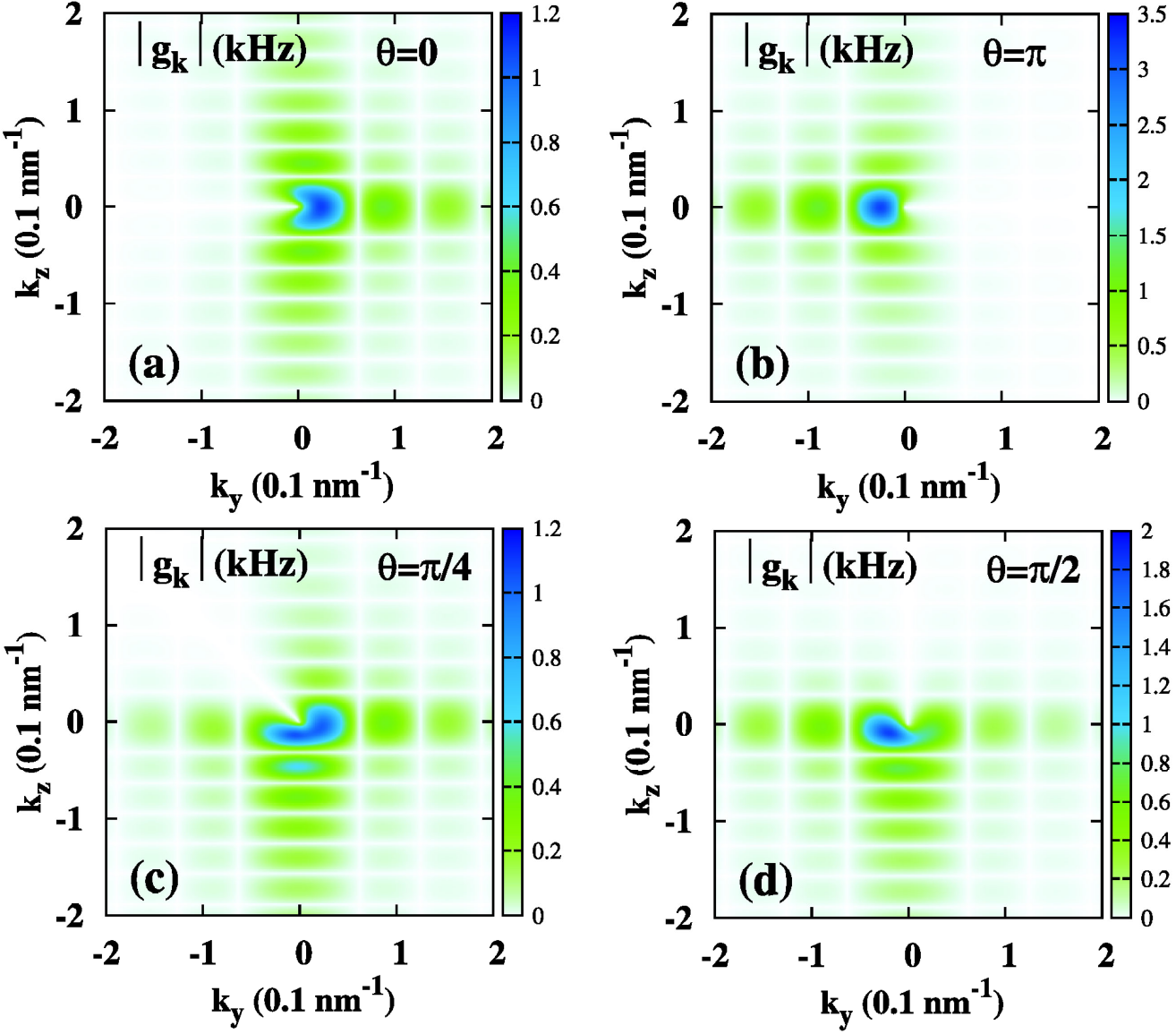} 
\caption{Coupling constants $g_{\mathbf{k}}$ between Kittel magnons in the nanomagnet and propagating magnons with wave vector $\mathbf{k}$ in the film when biased by the applied magnetic field along different directions $\theta=\{0,\pi,\pi/4,\pi/2\}$. The equilibrium magnetization of the nanomagnet is pinned along the $\hat{\bf z}$-direction by the shape anisotropy. The soft film magnetization follows here the applied field $\mathbf{H}_0$.} 
\label{gk_1}
\end{figure}

The interaction of magnons in the film with the stray field of the Kittel magnon in the nanomagnet is modified by an anisotropy to 
\begin{align}
&g_{\mathbf{k}}   =-4\mu_{0}\gamma\sqrt{M_{s}\tilde{M}_{s}}\frac{\left(
1-e^{-kd}\right)  \left(  1-e^{-ks}\right)  }{k^{3}k_{y}k_{z}}\sin\left(
\frac{k_{y}w}{2}\right)\nonumber\\&\times\sin\left(  \frac{k_{z}l}{2}\right)  \left(  \mathcal{M}_{x},\mathcal{M}_{y}\right)  \left(
\begin{array}
[c]{cc}%
k^{2} & -ik\kappa_{1}\\
ik\kappa_{2} & \kappa_{1}\kappa_{2}%
\end{array}
\right)  \left(
\begin{array}
[c]{c}%
\tilde{\mathcal{M}}_{\tilde{x}}^{\ast}\\
\tilde{\mathcal{M}}_{\tilde{y}}^{\ast}%
\end{array}
\right),
\end{align}
where $\kappa_{1}=k_{y}\cos{\theta}-k_{z}\sin{\theta}$,
$\kappa_{2}=k_{z}\sin{\theta}-k_{y}\cos{\theta}$, $\tilde{\mathcal{M}}_{\tilde{x}}=-1/(2\xi\sqrt{lwd})$, and $\tilde{\mathcal{M}}_{\tilde{y}}=-i\xi/(2\sqrt{lwd})$. Figure~\ref{gk_1} plots the dependence of the coupling constant $g_{\mathbf{k}}$ on the wave numbers $k_{y}$ and $k_{z}$ for the nanomagnet in Fig.~\ref{spin_1}. The YIG film with magnetization $\mu_{0}{M}_{s}=0.177$ is $s=10~\mathrm{nm}$ thick.  

\begin{figure}[tbh]
\centering
\includegraphics[width=0.48\textwidth]{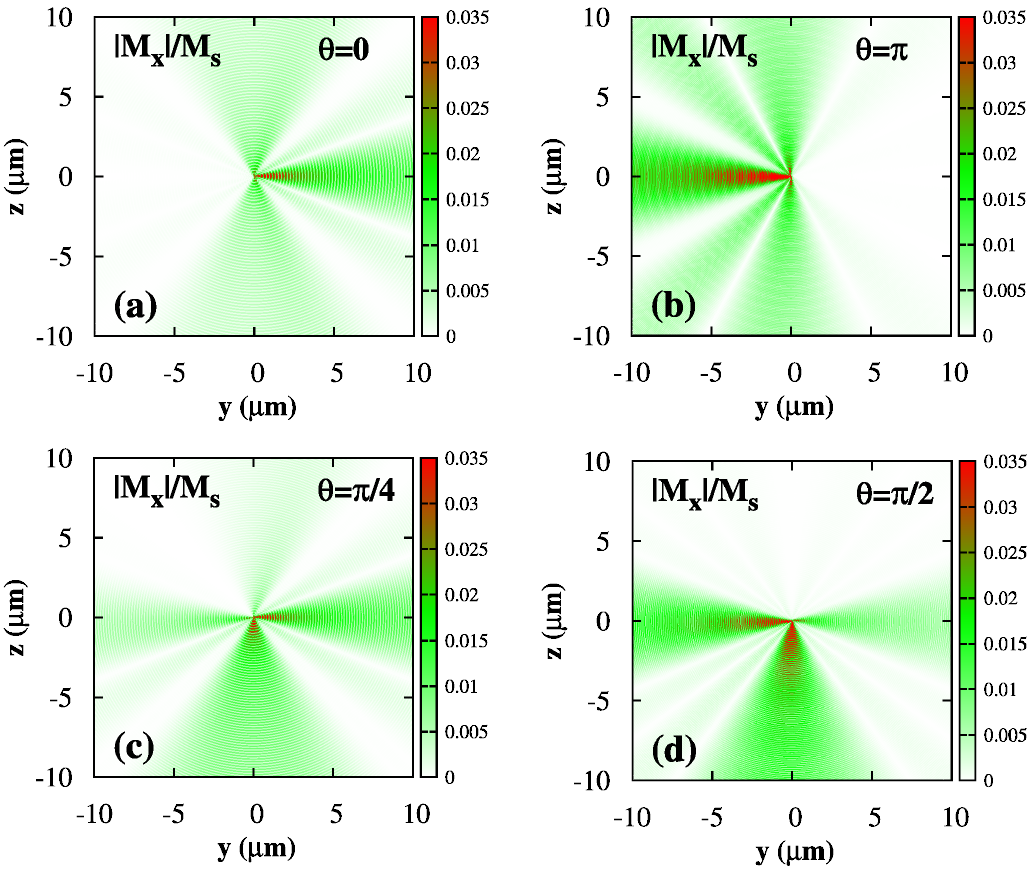} 
\caption{Spatial distribution of the magnetic  fluctuations in a YIG film below a resonantly excited CoFeB nanomagnet with equilibrium magnetization pinned along the $\hat{\bf z}$-direction  under an in-plane applied magnetic field of 50~mT in
different directions  $\theta=\{0,\pi,\pi
/4,\pi/2\}$ . }
\label{excitation_cubic}
\end{figure}

Figure~\ref{excitation_cubic} are plots of the
excited magnetization texture in a thin YIG film of $s=10~\mathrm{nm}$ below the CoFeB nanomagnet with dimensions $\{w,l,d\}=\{100,200,30\}~\mathrm{nm}$ and excitation amplitude
$\langle\hat{\beta}\rangle=3\times10^{5}$ or $\tilde{M}_{x}/\tilde{M}_{s}\approx0.04$, 
where $\mu_{0}{M}_{s}=0.177$~T, the exchange stiffness
$\alpha_{\mathrm{ex}}=3\times10^{-16}~\mathrm{m}^{2}$, the
Gilbert damping constant ${\alpha}_{G}=10^{-4}$, $\gamma=1.82\times
10^{11}~\mathrm{s^{-1}\cdot T^{-1}}$, and  $\mu_{0}\tilde{M}_{s}%
=1.6$~T. We find again an anisotropic ``lighthouse" distribution of the emitted magnons, i.e.,  beams that can be steered by the direction of YIG's (soft) equilibrium
magnetization. Comparing with the results for a disk-shaped magnet in Sec.~\ref{isotropic_II}, we conclude that the shape anisotropy significantly affects the angular distribution of the emitted magnons.

\section{Routing on-chip spin information}

\label{Section_V}

Long-distance entanglement~\cite{Einstein} enables the scalability of quantum  processors~\cite{qubitsscale}. In this context, magnons are intensely studied by theory~\cite{MasayaPRX,NeumanPRL,FlebusPRB,TrifunovicPRX,TrifunovicPRX2013} and experiments~\cite{Fukamicoupling}. For convenience, researchers focus mostly on configurations with one relevant spatial variable.  Here we address an on-chip controllable and long-distance coherent coupling of spin information stored in two distant nanomagnets by the exchange of virtual magnons. For quantum applications, diamond NV-centers~\cite{MasayaPRX,TrifunovicPRX2013,Candido2021} have advantages over nanomagnets, representing a two-level system with low damping.  The coupling of NV centers by magnon exchange ~\cite{Fukamicoupling} requires control of the distance to the magnetic film and the spin direction of single NV centers, and the coupling strength is weak. Here we focus on a pair of magnetic dots that can be fabricated and controlled relatively easily with a much stronger coupling in a single device.

\begin{figure}[tbh]
\centering
\includegraphics[width=1.0\linewidth]{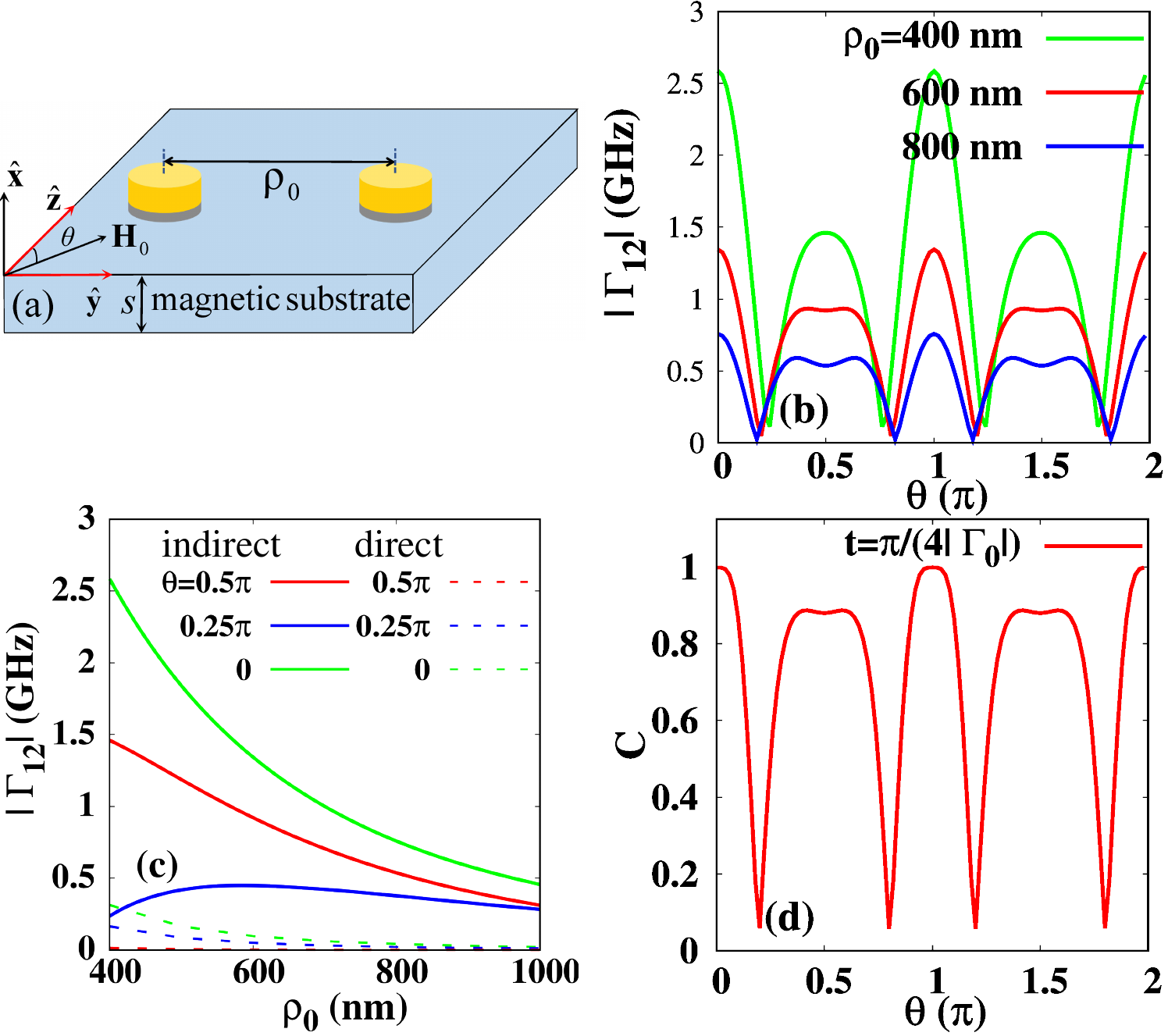}
\caption{Coupling and entanglement of two nanomagnets mediated by spin waves in a magnetic film. (a) illustrates the
configuration.  (b) shows the magnetization direction dependence of the coupling constant $\Gamma_{12}$ at $\rho_0=\{400,600,800\}~\mathrm{nm}$. (c) compares the indirect (solid curve) and direct (dashed curve) interactions as a function of the distance between nanomagnets for different magnetization directions. (d) shows a density plot of the entanglement as a function of $\theta$ with fixed time $t=\pi/(4|\Gamma_{0}|)$ at $\rho_0=600~\mathrm{nm}$, where $|\Gamma_{0}|$ is the maximal value of the coupling $\Gamma_{12}$. }
\label{angle_11}%
\end{figure}

We illustrate the physics at the hand of two identical disk-shaped nanomagnets at a distance $\rho_0$ on top of the magnetic film at
$(y,z)=(-\rho_0/2,0)$ and $(\rho_0/2,0)$, as illustrated in Fig.~\ref{angle_11}(a). The magnetic field and all the magnetizations point in the same direction.  
In contrast to the ``magnon trap"~\cite{magnon_trap} in which we considered the dissipative regime of exciting real spin waves, we focus here on resonance frequencies $\Omega$ that lie below the magnon band gap. 
We can then trace out the virtual magnons in the film to obtain an effective interaction Hamiltonian 
\[
\mathcal{H}_{\mathrm{eff}}=\hbar\hat{\mathcal{M}}^{\dagger}\left(
\begin{matrix}
{\Omega}+\Gamma_{11} & \Gamma_{12}\\
\Gamma_{21} & \tilde{\Omega}+\Gamma_{22}%
\end{matrix}
\right)  \hat{\mathcal{M}},
\]
where $\hat{\mathcal{M}}\equiv(\hat{\beta}_{1},\hat{\beta}_{2})^{T}$ collects the magnon operators for the two nanomagnets. The virtual magnons in the films red-shift the FMR frequencies by
$\Gamma_{11}=\Gamma_{22} = \sum_{\bf k}|{g{({\bf k},\theta)}}|^2/(\Omega-\tilde{\omega}_k)$,
and induce an effective coherent coupling $\Gamma_{12}(\rho_0,\theta)=\sum_{\mathbf{k}}{\left\vert {g{(\mathbf{k},\theta)}}\right\vert ^{2}e^{i\mathbf{k}\cdot{{\pmb \rho}_0}}}/({\Omega-\tilde{\omega}_{k}})\approx \Gamma_{21}^{\ast}$. When the Gilbert damping in the film  ${\alpha}_G$  is much larger than that of the nanomagnet $\tilde{\alpha}_G$,  we may describe dissipation by the complex dispersion relation $\tilde{\omega}_k=\mu_{0}\gamma(H_{0}+{\alpha}_{\mathrm{ex}}{M}_{s}k^{2})(1-i{\alpha}_G)$. 

Figure~\ref{angle_11}(b) illustrates the dependence of $\Gamma_{12}$ on the angle of an applied field with strength $\mu_{0}H_{0}=0.1$~T at constant distance between two equal YIG disks of dimensions $\{w,d\}=\{100,180\}$~nm on top of a thin CoFeB film of thickness $s=10~\mathrm{nm}$.  The magnons mediate an interaction over \(\mu\)m that strongly dominates the direct dipolar interaction (dashed line) addressed in the Appendix~\ref{appendix} [Fig.~\ref{angle_11}(c)]. Here we use the material parameters from Fig.~\ref{excitation_11}(d) with magnon frequencies in the film above that of the nanomagnet to ensure that the magnon exchange is not dissipative.

The exchange stiffness of CoFeB $\alpha_{\mathrm{ex}}=8\times
10^{-17}~\mathrm{m}^{2}$~\cite{Mook1,Pajda1} and its
Gilbert damping constant $\alpha_{G}=10^{-3}$~\cite{alphaG1}. The magnon band of a film with $\mu_0 M_s=1.6$~T~\cite{Heigl1}  $\omega(\mathbf{{k}})\geq\mu_{0}\gamma\sqrt{H_0(H_0+{M}_{s})}>\mu_{0}\gamma H_0=18.2~{\mathrm{GHz}}>\Omega=18.1~{\mathrm{GHz}}$ lies above the FMR frequency of the disks. 
Even though the chirality vanishes in the virtual excitation [see Fig.~\ref{excitation_11}(d)] the coupling $\Gamma_{12}$ is still strongly 
 angle-dependent, with maxima in the GHz regime at \(\theta=\{0,\pi\}\) with large cooperativities~\cite{cavity} ${\cal C}=4|\Gamma_{12}|^2/(\tilde{\alpha}_G\Omega)^2\sim 10^6$.  The photonic SOC forbids magnon exchange for angles \(\theta=\{\pi/4,3\pi/4\}\) at which the coupling nearly vanishes.

The anisotropic indirect coupling $\Gamma_{12}$ generates a tunable entanglement between two distant nanomagnets, as in Fig.~\ref{angle_11}(d). We now consider the quantum dynamics of the system initialized to a state with a
single magnon in one nanomagnet denoted as $\ket{1,0}$. The quantum dynamics
of the two nanomagnets obey the quantum master equation
for the density matrix $\rho$ at zero temperature 
\begin{align}
{d\rho}/{dt}=-({i}/{\hbar})[\mathcal{H}_{\mathrm{eff}},\rho
]+\delta_{\beta}\mathcal{L}[\beta_{1}]+\delta_{\beta}\mathcal{L}[\beta_{2}],
\end{align}
where the Lindblad dissipation operator $\mathcal{L}[\beta]\equiv\beta
\rho\beta^{\dagger}-\{\beta^{\dagger}\beta,\rho\}/2$ represents the magnon
damping. The concurrence~\cite{Hill1, Concurrence,ZouPRB}
\begin{align}
    C(t,\theta)=2\left\vert \rho_{12}\right\vert =e^{-\delta_{\beta}%
t}|\sin\left[  2\Gamma_{12}(\theta)t\right]|
\end{align} 
measures the time-dependent entanglement that exponentially decays with a lifetime $1/\delta_{\beta}=1/(\tilde{\alpha}_G\Omega)\sim 500$~ns that is governed by the Gilbert damping of the nanomagnets. 
At a fixed angle $\theta$, the concurrence is maximal for $t_0\equiv \pi/(4|\Gamma_{12}|)$, at which the two nanomagnets form a fully delocalized Bell-state $\ket{\varphi}=(|1,0\rangle-i|0,1\rangle)/\sqrt{2}$ with fidelity: 
\[
\mathcal{F}=\Tr\big[ \rho(t_0)\ket{\varphi}\bra{\varphi}\big] =\exp[-\pi\tilde{\alpha}_\text{G}{\Omega}/(4{|\Gamma_{12}|})]\rightarrow 1.
\]
The entanglement of distant magnons can be observed by Wigner tomography~\cite{Quantum11}.

\section{Conclusion}

\label{Section_VI}

To conclude, we report a geometric SOC in evanescent vector fields: it is always normal to its propagation, and the associated chirality is always right-handed. In nanomagnetic particles on top of ultrathin magnetic films, this leads to a direction-dependent excitation of magnons. The magnon-mediated coupling between two or more nanomagnets on top of a magnetic film can be strong up to large distances and tuned by the relative magnetization directions. The controlled high cooperativities facilitate scalable magnon-based classical or quantum information processors. Entangling a two-dimensional lattice of magnetic dots (magnonic crystal) is the next challenge in the quest for spin-based quantum information processing. 

\begin{acknowledgments}
This work is financially supported by the National Key Research and Development Program of China under Grant No. 2023YFA1406600, the National Natural Science Foundation of China under Grant No. 12374109, and the startup grant of Huazhong University of Science and Technology. J.Z. acknowledges the support of the Georg H. Endress Foundation. The JSPS KAKENHI Grants No. 19H00645, 22H04965, and JP24H02231 support G.B. financially. We thank Mehrdad Elyasi and Eugene Kamenetskii for illuminating discussions. 
\end{acknowledgments}

\begin{appendix}

\section{Direct dipolar interaction between two nanomagnets}
\label{appendix}

Here we address the direct dipolar interaction between the Kittel magnons of two identical magnetic disks of radius $w$ and thickness $d$, centered at $(x,y,z)=(d/2,0,0)$ and $(d/2,\rho_0,0)$, as in Fig.~\ref{angle_11}(a). We introduce a local $\{\tilde{x},\tilde{y},\tilde{z}\}$-reference frame with ${\tilde{\bf z}}\parallel {\bf M}_s$ and ${\tilde{\bf x}}\parallel \hat{\bf n}$ for convenience.
The magnetic field $H_0\tilde{\bf z}$  is applied in the plane at an angle  $\theta$ from the original \textbf{z}-axis and aligns both magnetizations.

Adapting the results from the main text, the Fourier components of the dipolar magnetic field emitted by a Kittel mode of a disk centered at $(d/2,0,0)$ read for  $0<x<d$ 
\begin{align}
    &h_{x} (x,q_y,q_z)={\sqrt{q_y^2+q_z^2}}\nonumber\\
    &\times\Big[-\tilde{M}_x\sqrt{{q_y^2}+{q_z^2}}\left( e^{-\sqrt{q_y^2+q_z^2}x}+e^{\sqrt{q_y^2+q_z^2}(x-d)}\right)\tilde{\cal V}_{q_y,q_z}\nonumber\\
        &+(\tilde{M}_y iq_y+\tilde{M}_z iq_z)\tilde{\cal V}_{q_y,q_z}\left( e^{-\sqrt{q_y^2+q_z^2}x}-e^{\sqrt{q_y^2+q_z^2}(x-d)}\right)\Big],\nonumber\\
       &h_{y} (x,q_y,q_z)=iq_y\nonumber\\&\times \Big[\tilde{M}_x \sqrt{q_y^2+q_z^2}\left( e^{-\sqrt{q_y^2+q_z^2}x}-e^{\sqrt{q_y^2+q_z^2}(x-d)}\right)\tilde{\cal V}_{q_y,q_z}\nonumber\\
        &+(\tilde{M}_y iq_y+\tilde{M}_z iq_z)V_{q_y,q_z}(x) \Big],\nonumber\\
        &h_{z} (x,q_y,q_z)=iq_z\nonumber\\&\times\Big[\tilde{M}_x \sqrt{q_y^2+q_z^2}\left( e^{-\sqrt{q_y^2+q_z^2}x}-e^{\sqrt{q_y^2+q_z^2}(x-d)}\right)\tilde{\cal V}_{q_y,q_z}\nonumber\\
        &+(\tilde{M}_y iq_y+\tilde{M}_z iq_z)V_{q_y,q_z}(x) \Big].
\end{align}
Here $\tilde{M}_x=-\sqrt{2\tilde{M}_{s}\gamma\hbar}\tilde{\mathcal{M}}_{\tilde{x}}$, $\tilde{M}_y=-\sqrt{2\tilde{M}_{s}\gamma\hbar}\tilde{\mathcal{M}}_{\tilde{y}}\cos\theta$, and $\tilde{M}_z=\sqrt{2\tilde{M}_{s}\gamma\hbar}\tilde{\mathcal{M}}_{\tilde{y}}\sin\theta$ in terms of the normalized amplitudes of the Kittel modes $\tilde{\mathcal{M}}_{\tilde{x}}$ and $\tilde{\mathcal{M}}_{\tilde{y}}$. The form factors   $\tilde{\cal V}_{q_y,q_z}=\pi w J_{1}(w\sqrt{q_y^2+q_z^2})/(q_y^2+q_z^2)^{3/2}$ and $V_{q_y,q_z}(x)=\tilde{\cal V}_{q_y,q_z}\left(2-e^{-\sqrt{q_y^2+q_z^2}x}-e^{\sqrt{q_y^2+q_z^2}(x-d)}\right)$, where $J_1(x)$ is the first-order Bessel function of the first kind. 
At $x=d/2$: 
\begin{align}
    &h_{x} \left(\frac{d}{2},q_y,q_z\right)=2\bigg[-\tilde{M}_x\sqrt{{q_y^2}+{q_z^2}}\cosh\left(\frac{d}{2}\sqrt{q_y^2+q_z^2}\right)\tilde{\cal V}_{q_y,q_z}\nonumber\\&+(\tilde{M}_y iq_y+\tilde{M}_z iq_z)\tilde{\cal V}_{q_y,q_z}\sinh\left(\frac{d}{2}\sqrt{q_y^2+q_z^2}\right)\bigg]
   {\sqrt{q_y^2+q_z^2}},\nonumber\\
    &h_{y}\left(\frac{d}{2},q_y,q_z\right)=  \bigg[2\tilde{M}_x \sqrt{q_y^2+q_z^2}\sinh\left(\frac{d}{2}\sqrt{q_y^2+q_z^2}\right)\tilde{\cal V}_{q_y,q_z}\nonumber\\&+(\tilde{M}_y iq_y+\tilde{M}_z iq_z)V_{q_y,q_z}\left(\frac{d}{2}\right) \bigg]iq_y,\nonumber\\
    &h_{z} \left(\frac{d}{2},q_y,q_z\right)=\bigg[2\tilde{M}_x \sqrt{q_y^2+q_z^2}\sinh\left(\frac{d}{2}\sqrt{q_y^2+q_z^2}\right)\tilde{\cal V}_{q_y,q_z}\nonumber\\&+(\tilde{M}_y iq_y+\tilde{M}_z iq_z)V_{q_y,q_z}\left(\frac{d}{2}\right) \bigg]iq_z.
\end{align}

Placing the second nanomagnet at $(d/2,\rho_0,0)$, the interaction Hamiltonian of the Kittel magnons $\{\hat{\beta}_1,\hat{\beta}_2\}$ in the two nanomagnets in vacuum reads 
\begin{align}
    \hat{H}_{c}=-\mu_{0}\int d\mathbf{r}\hat{\mathbf{h}%
}(\mathbf{r})\cdot\hat{{\mathbf{M}}}(\mathbf{r})=\hbar\Gamma^d_{12}
\hat{\beta}_1^{\dagger}\hat{\beta}_2+\mathrm{H.c.},
\end{align}
where the dipolar coupling constant 
\begin{align}
&\Gamma^d_{12} =-\frac{\mu_0\sqrt{2\tilde{M}_{s}\gamma}{\pi w^2d}}{\sqrt{\hbar}}\sum_{q_{y},q_{z}}e^{iq_{y}\rho_0}\bigg[\tilde{\mathcal{M}}_{\tilde{x}} h_{x}\left(\frac{d}{2},q_y,q_z\right)\nonumber\\&+\tilde{\mathcal{M}}_{\tilde{y}}\left( h_{y}\left(\frac{d}{2},q_y,q_z\right)\cos\theta + h_{z}\left(\frac{d}{2},q_y,q_z\right)\sin\theta\right) \bigg].
\end{align}
The direct dipolar coupling $\Gamma^d_{12}(\rho_0)$ between two nanomagnets is nearly proportional to $1/\rho_0^3$ in the far region with \(\rho_0 \gg \{d,w\}\).

\end{appendix}

\end{document}